\documentclass[runningheads, fullpage]{llncs}
\usepackage{microtype} % professional-grade typography with less need for hyphenation
\usepackage{graphicx} % support the \includegraphics command and options
\usepackage{url}
\usepackage{cite}
\usepackage{hyperref}
\newcommand{\Check}[1]{}

\newcommand{\VTNote}[1]{}

\newcommand{\commentOut}[1]{}

\usepackage{color}
\usepackage{soul}

\newif\ifblind
\blindfalse % Change this to enable/disable blind submission.
\title{The New South Wales iVote System:\\Security Failures and Verification Flaws\\in a Live Online Election}
\titlerunning{The New South Wales iVote System: Security Failures}

\ifblind\relax\else
\author{J. Alex Halderman\inst{1} \and Vanessa Teague\inst{2}\thanks{To whom correspondence should be addressed.}} % Funding moved to acknowledgments.
\authorrunning{Halderman and Teague}
\institute{University of Michigan\\\email{jhalderm@eecs.umich.edu}
\and University of Melbourne\\\email{vjteague@unimelb.edu.au}}
\fi

\begin{document}
\maketitle

\begin{abstract}
In the world's largest-ever deployment of online voting, the iVote Internet voting system was trusted for the return of 280,000 ballots in the 2015 state election in New South Wales, Australia.  During the election, we performed an independent security analysis of parts of the live iVote system and uncovered severe vulnerabilities that could be leveraged to manipulate votes, violate ballot privacy, and subvert the verification mechanism.  These vulnerabilities do not seem to have been detected by the election authorities before we disclosed them, despite a pre-election security review and despite the system having run in a live state election for five days.  One vulnerability, the result of including analytics software from an insecure external server, exposed some votes to complete compromise of privacy and integrity.  At least one parliamentary seat was decided by a margin much smaller than the number of votes taken while the system was vulnerable.  We also found protocol flaws, including vote verification that was itself susceptible to manipulation.   This incident underscores the difficulty of conducting secure elections online and carries lessons for voters, election officials, and the e-voting research community.
\end{abstract}

\section{Introduction}

Internet voting has rarely been used in significant elections for public office, due to numerous, well established security risks~\cite{NIST}, such as compromise of election servers, of voters' client devices, of the network in between, and of the voter authentication process.  To better understand how these risks can play out in real elections, we studied what may be the largest deployment of Internet voting to-date, the March 2015 state election in New South Wales, Australia.

In this election, voters had the option to use an online voting system called iVote, which was developed by e-voting vendor Scytl in partnership with the New South Wales Electoral Commission (NSWEC).  Prior to the election, NSWEC performed multiple security studies (e.g.~\cite{ivoteSec,ivoteThreats}), and officials publicly claimed that the vote was ``\textellipsis completely secret. It's fully encrypted and safeguarded, it can't be tampered with''~\cite{NSWECAssurance}. Over 280,000 votes were returned through iVote (about 5\% of the election total), exceeding the 70,090 Norwegian votes submitted online in 2013~\cite{nostats} and the 176,491 online votes in the 2015 Estonian election~\cite{eestats}.
\VTNote{At the risk of being a pedant, we don't actually know how many people voted in any of these cases, just how many votes were produced.  Not (necessarily) the same.}

While the election was going on, we performed an independent, uninvited security analysis of public portions of the iVote system.  We discovered critical security flaws that would allow a network-based attacker to perform downgrade-to-export attacks~\cite{dambh:freak,logJam}, defeat TLS, and inject malicious code into browsers during voting. We showed that an attacker could exploit these flaws to violate ballot privacy and steal votes.  We also identified several methods by which an attacker could defeat the verification mechanisms built into the iVote design.

After we reported these problems to authorities, NSWEC patched iVote to correct the network security flaws, but by this time the election had been running for five days and 66,000 votes had been cast on the vulnerable system.   
After the vulnerabilities were removed, we made our findings public %
\ifblind
[\emph{self-citations omitted}].
\else
in a  technical blog post on \emph{Freedom to Tinker}~\cite{FreedomToTinker}  and an essay for nontechnical readers in \emph{The  Conversation}~\cite{TheConversation}.
\fi

The election count is now complete~\cite{NSW2015LCResults}, with the final seat in the proportionally represented 
Legislative Council having come down to a margin of 3177 votes, a tiny fraction of the number of votes cast over iVote before it was patched.
To our knowledge, this is the first time enough votes to affect a parliamentary seat in a state election have been returned over an Internet voting system while it was demonstrably vulnerable to attacks that would allow external vote manipulation.  
While we do not know whether anyone exploited the opportunity for electoral fraud, we know the opportunity was there.

In this paper, we detail our security findings about iVote and draw broader lessons from this case study.  The iVote vulnerability reinforces findings of security problems in other proposed and fielded Internet voting systems, such as Washington, D.C.'s~\cite{dcFC} and Estonia's~\cite{estoniaCCS}, and it demonstrates once again that no amount of pre-election review can guarantee that such a system is secure.  These problems also highlight the brittleness of the web platform and TLS protocol---a fragility which may be incompatible with the intensive security requirements and time pressure of political elections.  

iVote's vulnerabilities should encourage skepticism of other Internet voting systems claimed to be verifiable.  Years of research on electronic methods of election verification are only just beginning to produce end-to-end verifiable voting systems appropriate for use in low-stakes, low-coercion elections \cite{adida2009electing}, or in government elections using a postal mail step~\cite{zagorski2013remotegrity}, or in the much easier case of supervised polling places \cite{vvotePaper, carback2010scantegrity, bell2013star}.  The iVote verification protocol ignores basic insights and techniques from that research, opting instead for a telephone-based vote reading service that substantially reduces voter privacy while providing only very limited assurances of integrity.  Furthermore, an election verification protocol, like any other security protocol, should not be  relied upon without an extensive period of public review; in the case of the iVote protocols, there was none.

Securing Internet voting requires overcoming some of the most difficult problems in computer security, and, with existing technology, even the smallest mistakes can undermine the integrity of the election result.  The experience in New South Wales is a real-world example demonstrating online voting security problems that many security researchers, including us, have warned about for many years.  We recommend that election officials refrain from conducting high-stakes elections online until there are fundamental security advances.

\section{iVote Background}

The iVote system was a complex interaction of many components, some managed by the NSWEC and some by other administrators.   Registration and voting could each be done by three different methods: by  telephone, over the Internet, or from a NSWEC computer in a polling place.  
There were four steps in using iVote:
\begin{enumerate}
\item The voter registered, received an 8-digit iVote ID, and chose a 6-digit PIN\@.
\item  The voter logged in to the voting server (or the telephone voting system) with her iVote ID and PIN, cast a vote, and received a 12-digit receipt number.  The vote was encrypted on the client, sent to the voting server, and forwarded to a separate verification service.
\item  Optionally, the voter telephoned the verification service, an  interactive voice response (IVR) system.  She entered her iVote ID, PIN, and receipt number and heard her vote read back.  This service stopped at the close of polls.
\item  Optionally, the voter visited an online receipt service to query whether any votes with her receipt number were included in the final count. No login was needed.  This service remained active after the close of polls.
\end{enumerate}

More details are described in the Security Implementation Statement~\cite{ivoteSec} and other reports published by NSWEC~\cite{ivoteReports}.  These include prose descriptions of the methods of encrypting and processing the vote.  The protocol evolved over several drafts, but all of them differ in some important respects from what the system actually did during the election (see Section~\ref{sec:auditAndVerification}).
% AH: Cutting this as unused detail.  Happy to undo if we need it.
\if0 These documents describe procedures for a ``decryption verification ceremony'', in which an auditor could reconcile the votes on the verification server with the main election data.  There are some procedures for removing duplicates and votes that were cast via a different channel, and for testing by running dummy votes through the process during the election and then removing them afterwards.\fi
No source code was made available for any of the server-side processes, including the main voting web server, verification server,  registration server, and receipt server.  Anyone could, of course, inspect the HTML and JavaScript delivered to the voter's browser.

\begin{figure}[t]
    \centering
    \includegraphics[width=\linewidth]{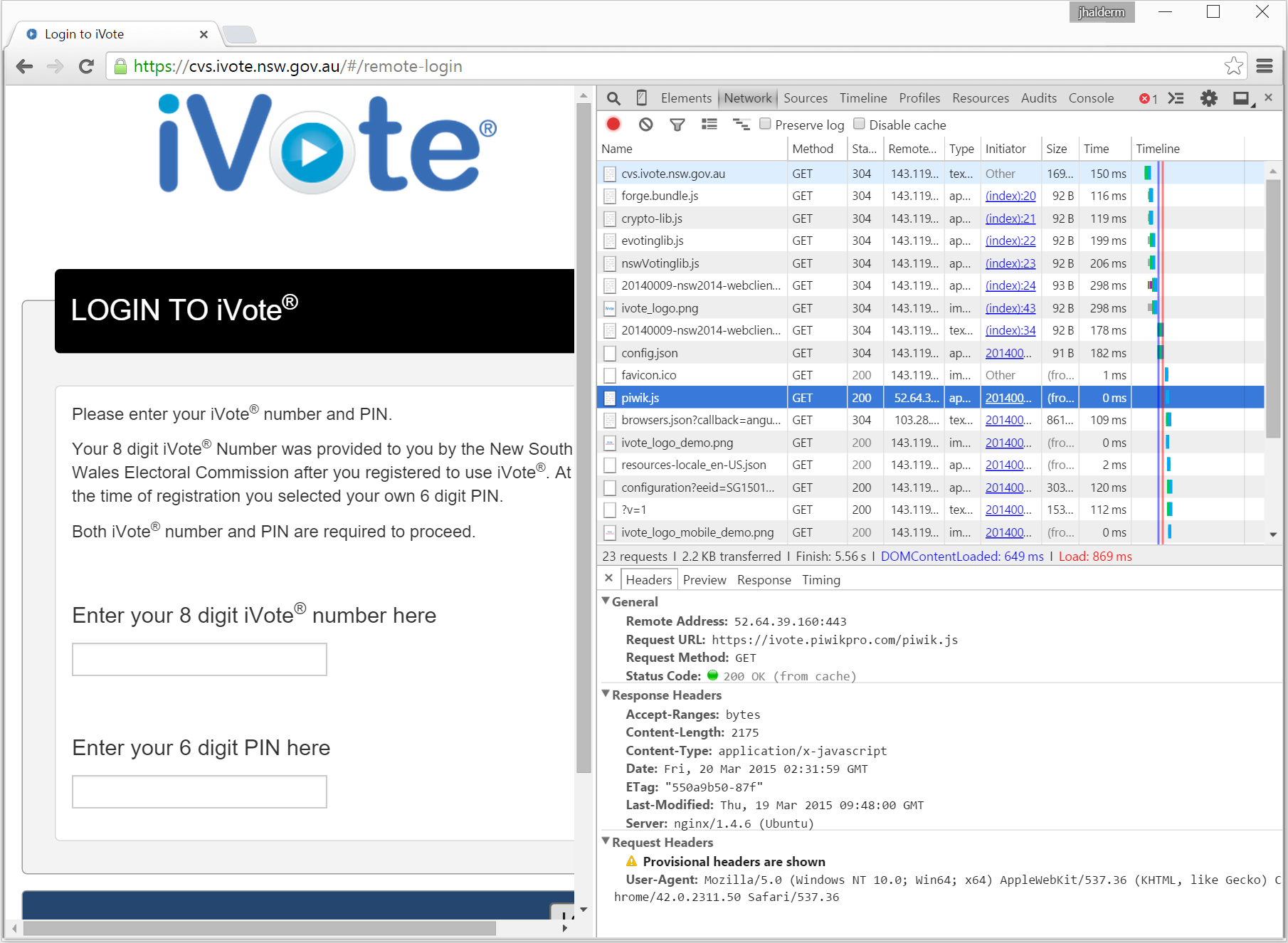}%ethertex: https://www.dropbox.com/s/5vhphx2o1tasroz/piwik.png?dl=1
    \caption{Like most web applications, iVote was made up of dozens of resources that were loaded in the background by the browser.  Using the Chrome Developer Tools, we could see that most of the iVote resources came from the ``core voting system'' server, \texttt{cvs.ivote.nsw.gov.au}, but one component, JavaScript for the Piwik analytics tool, was loaded from an external server, \texttt{ivote.piwikpro.com}.}
    \label{fig:piwikInIvote}
\end{figure}

In the 2015 state election, each voter could cast one vote for the Legislative Assembly and one for the Legislative Council.  Although iVote was officially reserved for the disabled and other eligible absentee voters, voters could qualify by self-certifying that they would be out of the state during election day~\cite{ivoteFAQ}.  %VT: cutting lines.
iVote opened to the public on the morning of March 16 and closed at 6~P.M. on March 28, the same time as other polls closed in the state election.  Officials reported that about 280,000 votes (5\% of all counted ballots) were cast  over iVote.

% AH: Cutting this as unnecessarily personal.  We can add back in for camera ready if you like!  : )
\if1
This was the second state election in which NSWEC has run Internet voting, having used an earlier version (provided by Everyone Counts) in 2011.  We have offered them over the years explanations of the general risks associated with Internet voting and specific analysis of problems in the protocols proposed for the 2015 state election.  Very little of this advice was acted upon.  When we made our recent findings public (after the vulnerability was removed) NSWEC responded by denying the seriousness of the problem, attacking us in the press and making a formal complaint to the University of Melbourne claiming a ``breach of [the university's] policy on freedom of speech.''  We would have much preferred to persuade NSWEC not to run  Internet voting, than to have discovered that 66,000 votes were cast while iVote was vulnerable to manipulation and privacy breach.
\fi

\section{Vulnerabilities in iVote}\label{sec:security}

Shortly after iVote voting opened, we began an independent security review of the publicly accessible components of the system.  Although election officials did not publish the source code, client-side portions of this code were necessarily delivered to voters' browsers.  Since we were not eligible voters, we did not proceed past the login screen of the voting web application, \texttt{https://cvs.ivote.nsw.gov.au}, but we did inspect the HTML, CSS, and JavaScript code that made up the application.  In addition, NSWEC made a practice version of the iVote system available to the public at \url{https://practise.ivote.nsw.gov.au}.  The practice site allowed anyone to log in using provided credentials and vote a mock ballot.  We confirmed that the practice system used substantially the same client-side code as the real election server and used it to perform further hands-on tests.

\begin{figure}[t]
    \centering
    \includegraphics[width=\linewidth]{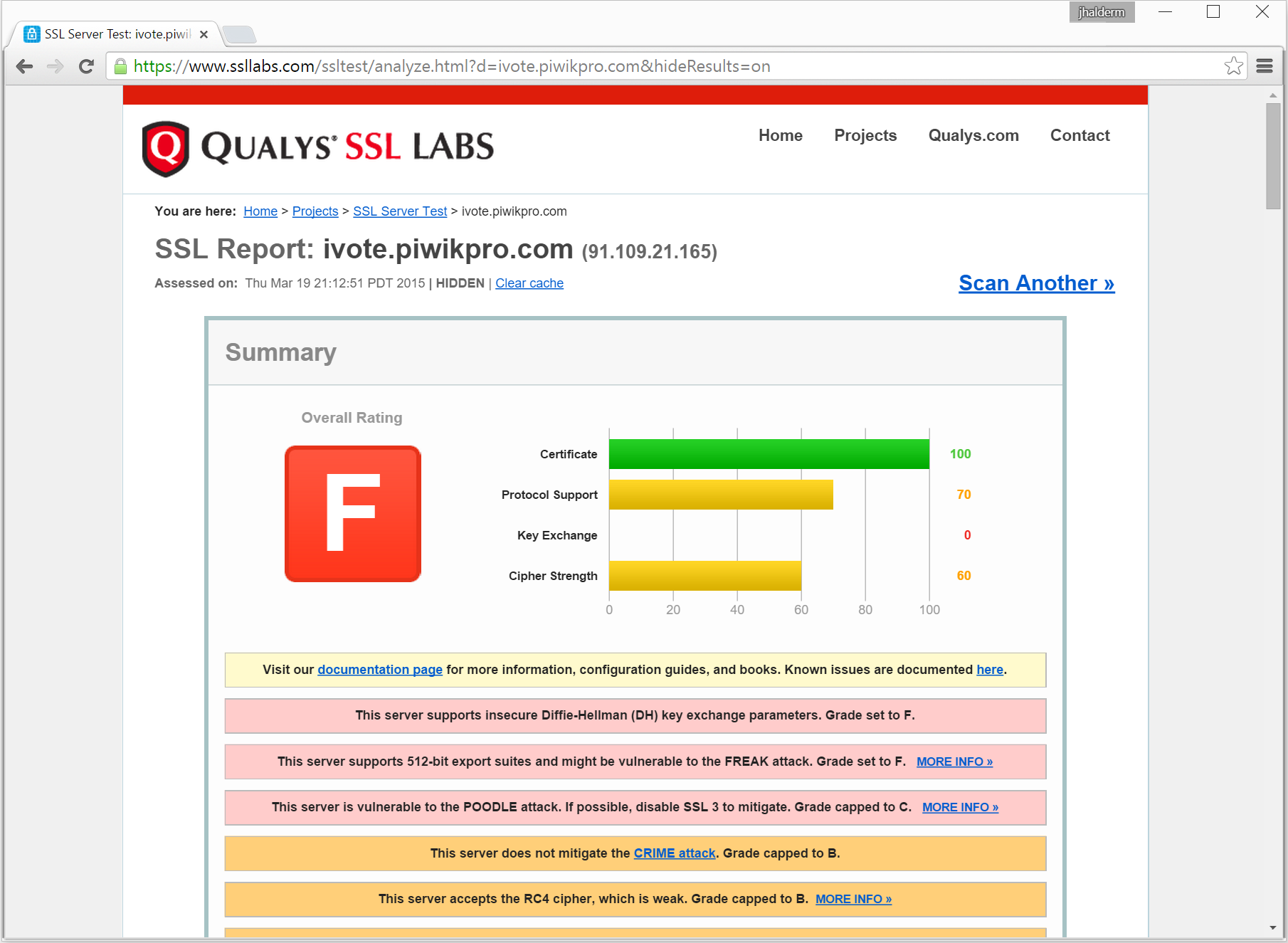}%ethertex: https://www.dropbox.com/s/rn9i4wh8rfk7uen/ssllabs-piwik.png?dl=1
    \caption{The \texttt{ivote.piwikpro.com} server scored an F on the Qualys SSL Labs tests.  Among other reported problems, the server used insecure Diffie-Hellman parameters, allowed 512-bit export cipher suites that are subject to the FREAK attack, and was vulnerable to the POODLE attack.  We showed that these problems would allow a man-in-the-middle attacker to inject vote-stealing code into the iVote application.\looseness=-1}
    \label{fig:FforPiwik}
\end{figure}

iVote is designed to deliver the web application using HTTPS\@.  
This is intended to prevent an adversary from modifying or replacing the code in transit to the user's web browser.  (The system uses other layers of cryptography on top of HTTPS to protect the actual ballot and submitted votes.)
However, not all HTTPS servers are secure---there are many configuration and operational details that system administrators need to get right in order to ensure that the protocol provides the desired security guarantees.  We tested the security of the main iVote HTTPS server using a standard tool, the Qualys SSL Labs SSL Test.  The results indicated that the server configuration complied with current best practices and was secure against known vulnerabilities.

However, a closer analysis of the structure of the iVote application showed that one of the resources loaded by the site came from an external web server.  When the voter loads the iVote site, the site imports and executes JavaScript for a third-party analysis tool called Piwik.  As shown in Figure~\ref{fig:piwikInIvote}, this code is loaded from a URL at the third-party server \url{https://ivote.piwikpro.com}.  When we tested the SSL configuration of this site, we found that it was extremely poor---scoring an `F' grade in the SSL Labs test, as shown in Figure~\ref{fig:FforPiwik}.  Among a variety of other security problems, the server supported 512-bit ``export-grade'' ciphersuites for both RSA and ephemeral Diffie-Hellman key exchange.  As we will show, this weak configuration allowed multiple ways for an attacker to bypass the security provided by HTTPS and inject malicious code into the user's iVote session without triggering any browser security warnings.

\subsection{Vulnerability to the FREAK attack}

The FREAK attack~\cite{beurdouche2015messy, dambh:freak}, short for Factoring {RSA} Export Keys, is a TLS vulnerability that was publicly disclosed on March 3, 2015, less than two weeks before the start of the election.  The Piwik server's configuration problems made it vulnerable to FREAK, and a network-based man-in-the-middle attacker could exploit the attack against the Piwik server in order to compromise iVote.  

As the name implies, FREAK exploits the weakness of 512-bit ``export-grade'' RSA keys that are supported by the TLS protocol as a legacy feature of 1990s era U.S. cryptographic export restrictions.  If a server supported export-grade RSA---as did \url{ivote.piwikpro.com}---an attacker could fool many popular browsers into using this reduced-strength cryptography, obtain the RSA private key by factoring the 512-bit public key, and manipulate the contents of the connection.

\newcommand{\tlshello}{\textsf{\small CLIENT\_HELLO}}

The attack begins by intercepting the browser's TLS \tlshello\ message and sending a substitute message to the server declaring that the browser wishes to use export-grade RSA\@.  In export-grade RSA modes, the server sends a 512-bit ``temporary'' RSA public key to the client and signs this key, together with a nonce chosen by the client, using the public key from its normal X.509 certificate.  The client verifies that there is a valid chain of certificates from the server's X.509 certificate to a trusted root certificate authority, then uses the temporary RSA key to encrypt session key material that will be used to secure the remainder of the connection. The FREAK attack exploits a mistake in the way browsers process the server's message containing this temporary key.  Several widely used TLS implementations would accept a temporary export-grade RSA key \emph{even if the client did not ask for it}.  This allows the attacker to downgrade a connection requesting normal RSA encryption to much weaker export-grade RSA\@.

The main challenge for the attacker is to convince the voter's browser that he really is \url{ivote.piwikpro.com}.  To do this, he needs the server's signature on the client's TLS nonce and an RSA public key that he knows the private key for.  Assume for now that Piwik always uses the same 512-bit key.  Nadia Heninger has shown that it is possible to factor 512-bit RSA keys using open-source software and Amazon EC2 in about 7 hours at a cost of about \$100~\cite{HeningerFactoring}.  Once the attacker has factored the key, he can intercept the user's connection, note the client's nonce, and make a request to the real Piwik server with the same nonce---in effect, using it as a signature oracle.  He can send the resulting signature on the RSA key as part of the connection to the voter's browser, which will see the key as valid and use it to encrypt its session key material.  Since the attacker has factored the key, he can decrypt this key material and impersonate the Piwik server for the rest of the connection.

One complication is that the Piwik server, unlike many TLS implementations, periodically rotated its temporary key.  In our tests, we saw the key change approximately every hour---too frequently to apply simple factoring methods available to us.  However, we found that we could force the Piwik server to use the same temporary RSA key for much longer periods by maintaining a long-lived TLS connection and repeatedly invoking client-initiated renegotiation.  Each renegotiation can use a different client nonce, so, by using this method, we could use the Piwik server as a signature oracle to attack as many clients as we wanted and use the same key for as long as this connection stayed open.

In tests, we were able to sustain the connection for 17--21 hours, and, with Heninger's assistance, we factored the temporary RSA key from one such session.  An attacker could start such a connection, spend the first 7 hours factoring the key, and then attack an unlimited number of voters' TLS connections for the remainder of the connection lifetime.  By making multiple such connections in a staggered fashion, the attacker could have continuously attacked iVote users for the duration of the connection at a cost of about \$100 per 12-hour period.

Many popular browsers were vulnerable to FREAK, including Internet Explorer, Safari, and Chrome for Mac OS and Android~\cite{dambh:freak}.  Although patches were released for most browsers around March 10, iVote voting opened on March  16, and many users likely had not applied the relevant patches.

\subsection{Vulnerability to the Logjam attack}

The {\tt ivote.piwikpro.com} server was also vulnerable to an even more powerful downgrade-to-export attack that affected \emph{all} popular browsers: the Logjam attack~\cite{logJam}, which was publicly disclosed on May 20, 2015.  We knew about this flaw during the election because one of us was part of the team that developed the attack, but we could not talk about it publicly because responsible disclosure to the browser-makers was still ongoing.  In other words, we had a zero-day TLS vulnerability that would have allowed us to attack any voter's iVote session.

Logjam is reminiscent of the FREAK attack, but it affects ephemeral Diffie-Hellman (DHE) ciphersuites rather than RSA ciphersuites, and it is made possible by a flaw in the TLS protocol rather than a client-side implementation error.  If a server supports export-grade Diffie-Hellman with parameters that an attacker can break, a man-in-the-middle can force browsers to use it, obtain the session keys, and intercept or arbitrarily change the contents of the connection~\cite{logJam}.

In Diffie-Hellman, two public parameters, a prime $p$ and a group generator $g$, are used to compute a public key $y$ from a secret key $x$ as $y = g^x \bmod p$.  An attacker can breach the security by computing the discrete logarithm of $y$ to recover $x$.  Although computing one discrete log is harder than factoring one RSA key of equivalent parameter size, a large part of the discrete log computation can be reused for all connections that use the same $p$~\cite{logJam}.

The Piwik server supported export Diffie-Hellman using a fixed 512-bit $p$:%
\if0\footnote{Based on Internet-wide measurements described in \cite{logJam}, this is the 40th most common 512-bit $p$ used for export-grade DHE\@.}\fi
\begin{quote}
\texttt{\small 
a705d4b834119d78e434e47be531ae602209c4810fa3baca2b781d49f847bc27\\
7681d93375522e41aae5de77d86d124852951be54145c9417f603ea96e5024b7}
\end{quote}
The team that developed the Logjam attack used open-source software to perform the precomputation step for three other common 512-bit values of $p$, each of which took about a week of wall-clock time using idle cycles on a cluster~\cite{logJam}.  Following precomputation, they could break individual key exchanges based on those values in about 90 seconds using a single 24-core machine.  The same kind of attack would be possible against Piwik's $p$, and would allow us to effectively attack all iVote sessions from any browser by paying a fixed up-front cost for the precomputation.  In that case, since the browser connects to Piwik in the background, the 90-second delay to compute the session key would not be noticeable by the voter.

\subsection{Proof-of-concept, exploitability, and responsible disclosure}

\begin{figure}[t]
    \centering
    \includegraphics[width=0.9\linewidth]{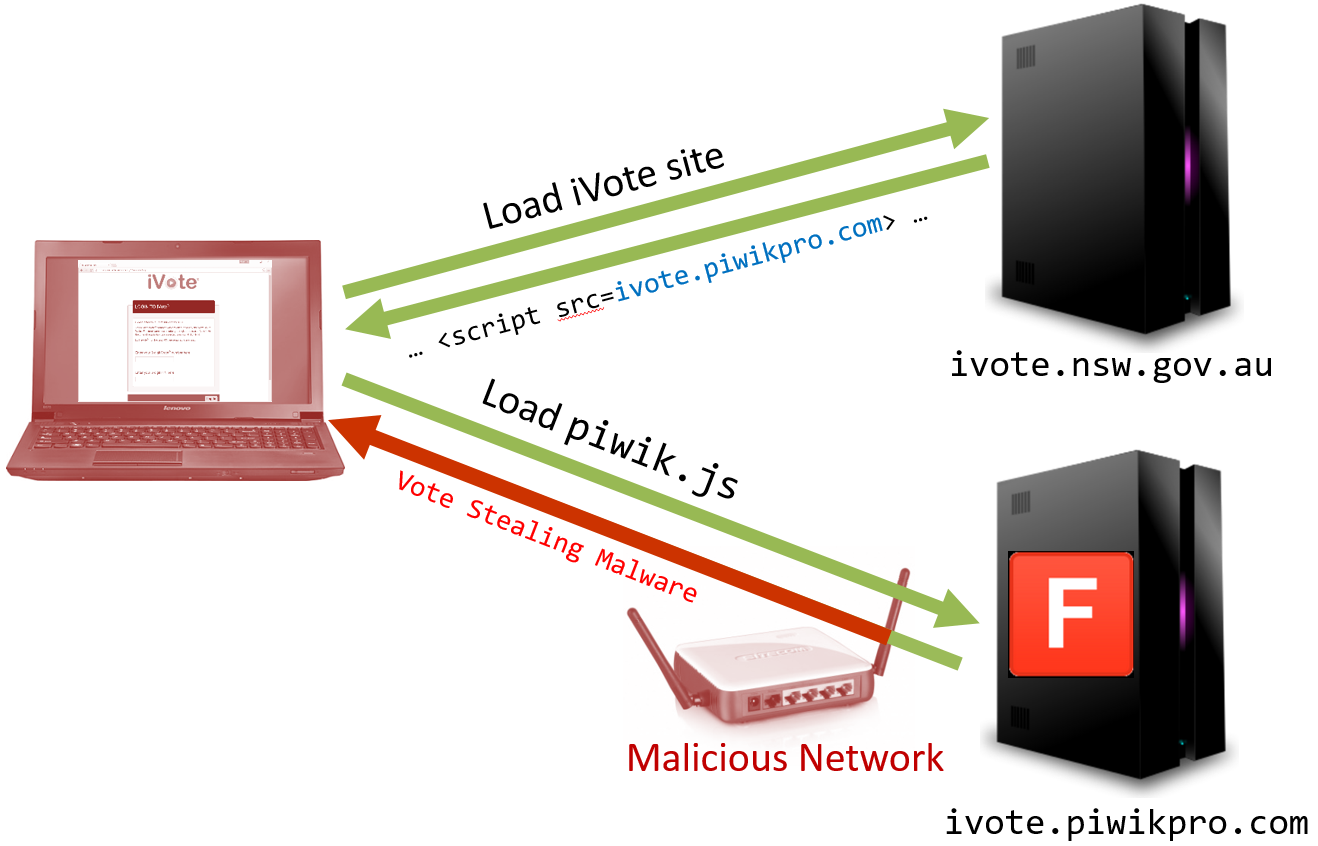}%ethertex: https://www.dropbox.com/s/zwx45lw14wox350/malware-diagram.png?dl=1
    \caption{Although the NSW web server used a secure HTTPS configuration to deliver the iVote application, the app subsequently loaded additional JavaScript from an insecure external server, \texttt{ivote.piwikpro.com}.  An attacker who intercepted connections between the voter's browser and the PiwikPro server could tamper with this JavaScript to inject arbitrary malicious code into the iVote application.\looseness=-1}
    \label{fig:malwareDiagram}
\end{figure}

We developed a proof-of-concept demonstration to show how an attacker could leverage the FREAK or Logjam vulnerability to manipulate the iVote system.  Following the scheme in Figure~\ref{fig:malwareDiagram}, this attack exploited the vulnerabilities in the Piwik server to replace the code loaded from \url{ivote.piwikpro.com} with malicious JavaScript.  Since this code was executed in the context of the user's iVote session, it could arbitrarily change the operation of the iVote web application.

\begin{figure}[t]
    \centering
    \includegraphics[width=0.49\textwidth]{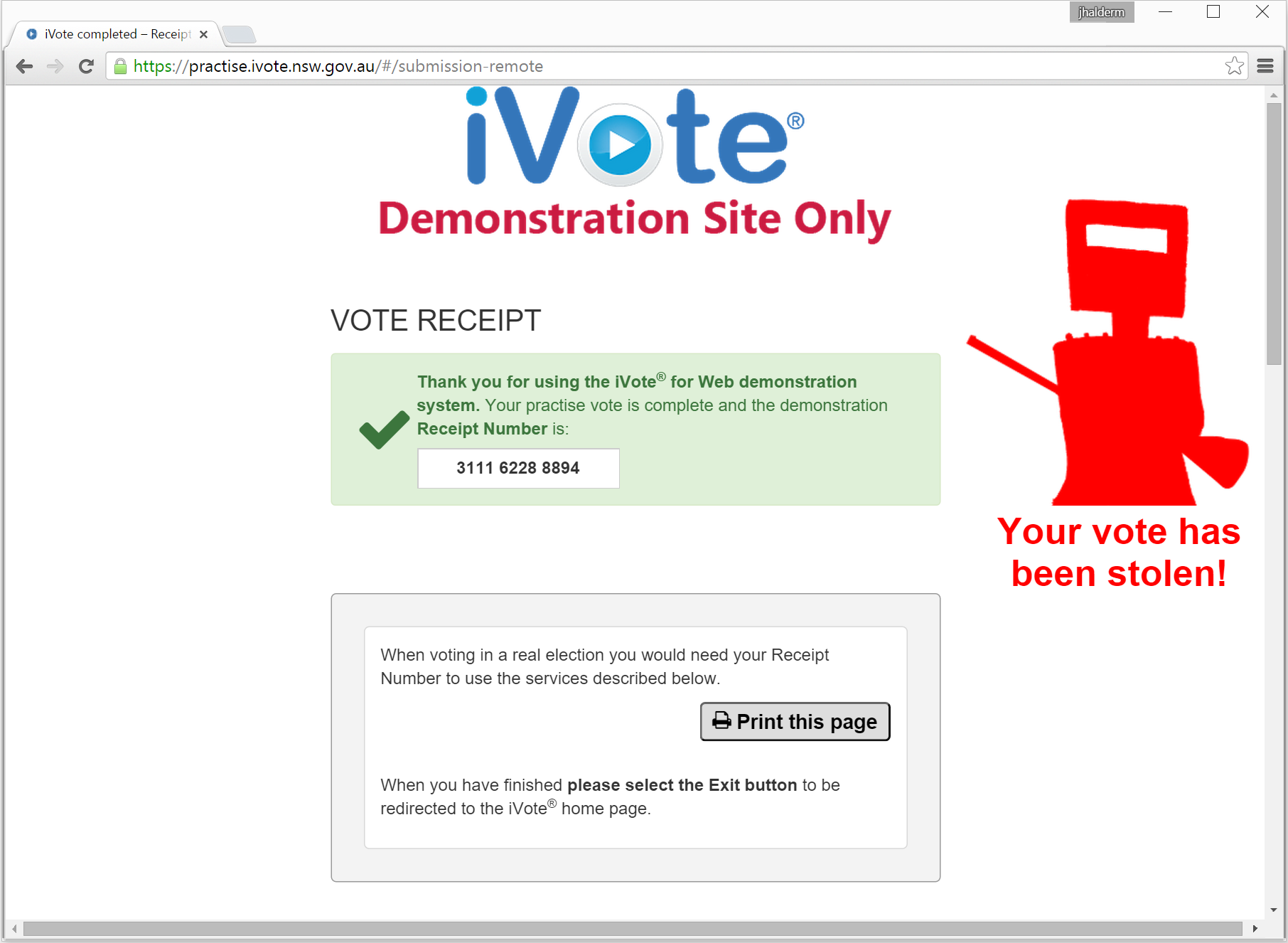}%ethertex: https://www.dropbox.com/s/tfpdm7tmlzus82u/receipt.png?dl=1    
    \hfill
    \includegraphics[width=0.49\textwidth]{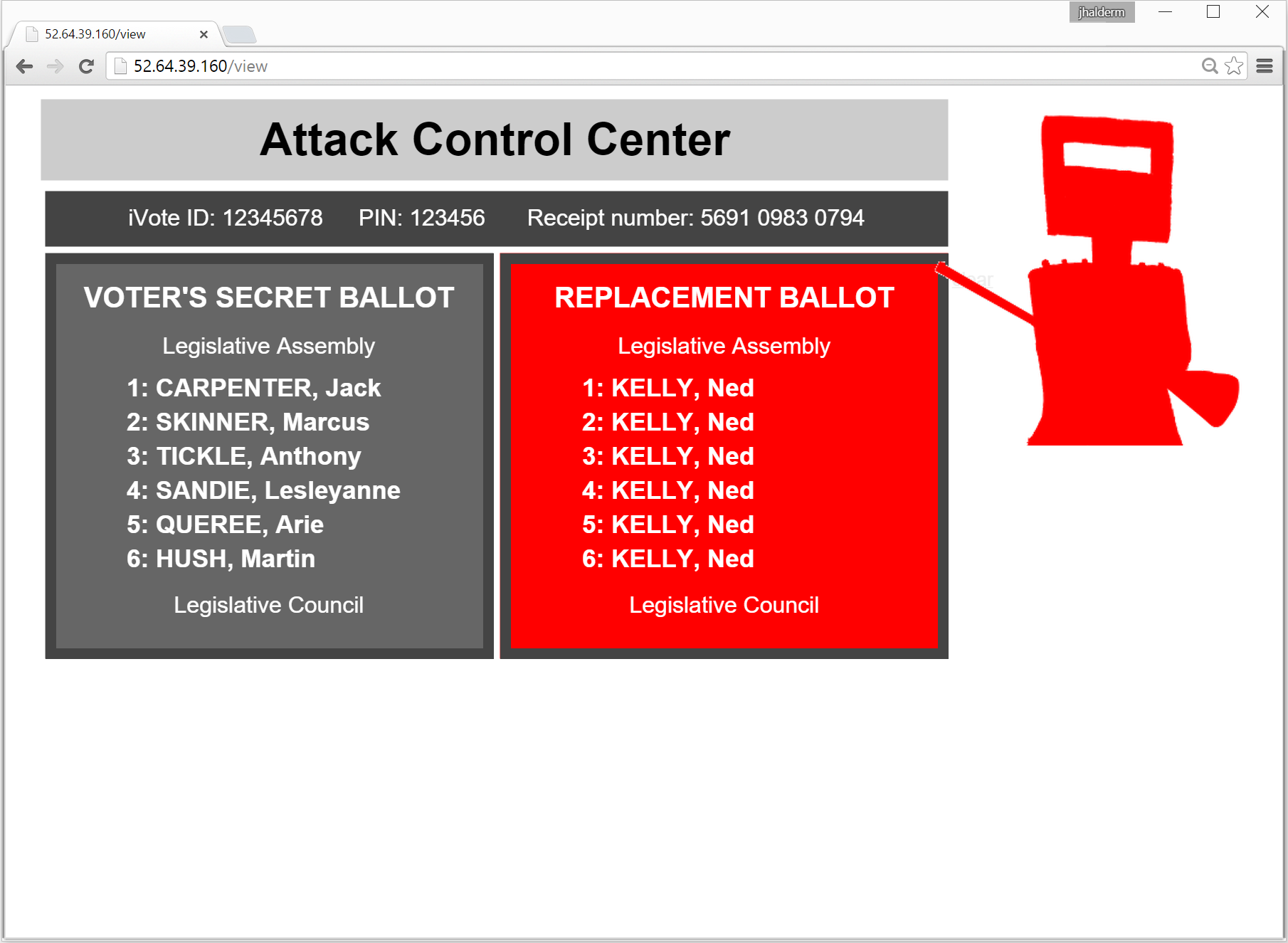}%ethertex: https://www.dropbox.com/s/bptwdb612es7odv/attacker.png?dl=1
    \caption{As a proof of concept, we showed that we could exploit the FREAK attack against iVote to inject malicious code that would surreptitiously manipulate the voter's choices (\emph{left}\/) and report them to a command-and-control server (\emph{right}\/).  Our mock attacker's symbol invokes Ned Kelly, an iconic Australian outlaw.\looseness=-1}
        \label{fig:ned}
\end{figure}

The demonstration malicious code we injected hooked into key parts of the iVote client code.  iVote used Angular{JS} to run a series of worker JavaScript threads which implemented cryptographic operations.  Crucial election data, including the contents of the vote, were passed between these workers as messages.  Our code intercepted these messages to change the intended vote to a different vote as it is passed to the worker script that performed the encryption.  This changed the vote that was sent to the iVote server.  Our code also exposed the vote that the voter intended to cast and sent it, along with the voter's authentication credentials, to a command-and-control server operated by the attacker.  Screenshots from our demonstration are in Figure~\ref{fig:ned}.

To exploit these attack against iVote, the attacker would need the ability to intercept and manipulate connections from the voter's browser destined for the Piwik server.  (Such man-in-the-middle attacks are, of course, one of the main threats that HTTPS is intended to guard against.)  Criminal attackers have many well documented ways to achieve this.  It could be done using client-side malware (including functions of widespread pre-existing botnets~\cite{dnsbotnet,MooseMalware}), by compromising insecure WiFi access points, by poisoning ISP DNS caches to redirect the traffic to an attacker-controlled IP address~\cite{dnspoisoning}, by attacking vulnerable routers or links along the path to the server, or by redirecting packets by hijacking BGP prefixes~\cite{bgphijacking}, to give just a few examples.  These attacks are especially practical in an election scenario, because the attacker can be highly opportunistic---he does not care which NSW voters he compromises, so he can target any insecure hosts or infrastructure in the entire state.  In addition to large scale criminal fraud, many individuals and employers have legitimate administrator privileges on home or workplace networks that others might use for voting, and could abuse these privileges to target votes.

Since we (of course) would not attempt to steal actual votes, we tested our demonstration attack only on our own votes, cast only on the iVote practice system, which was identical in all relevant respects to the real voting system.  After confirming that the attack was possible, we notified the Australian CERT of the vulnerabilities around 2~P.M. on  Friday, March 20.  CERT took responsibility for notifying the NSW Electoral Commission, which fixed the problem around midday on Saturday, March 21, by modifying the iVote server configuration to disable Piwik.  By then, about 66,000 votes had already been cast.  We cannot  know with certainty whether any real iVote votes were attacked; however, the final Legislative Council margin of 3177 votes represented less than 5\% of the votes cast over iVote while the server was vulnerable.

\section{Circumventing Verification}\label{sec:verification}

Vote manipulation attacks should be detectable with some probability by the verification mechanism.  However, the verification mechanism itself suffers from a number of straightforward circumventions and at least one important protocol flaw.\looseness=-1

\subsection{Simple verification avoidance}\label{subsec:simpleAvoidance}

The telephone-based verification scheme is easily sidestepped for last-minute votes because it shuts down at the close of polls.  
%(This closure seems to have been implemented in recognition of the threats to 
%vote privacy associated with the verification service's simultaneous knowledge
% of the voter's iVote ID, vote contents, and possibly telephone number.)  
So an attacker could confidently modify votes that were cast immediately before the  deadline, knowing that they could not be verified.  A malicious client (or server) could slow down near the end of polling to exacerbate this problem.

Voters are told how to verify by the same website they use to vote,  so the attacker could use the man-in-the-middle methods we describe above to direct the voter to a fake verification phone number that would read back the voter's intended choices.  Thanks to modern VoIP technology, setting up an automated phone system is simply a matter of software.

Even more simply, the attacker could delay submitting the vote and showing the receipt number for a few seconds, in hopes that the voter does not intend to verify and simply leaves the website.  (Perhaps the site could show a  progress bar in place of the number.)  If the voter navigates away, there will be no chance to verify, and the attacker can confidently submit a fraudulent vote.   Otherwise, the attacker can give up, submit the genuine vote,  and display the receipt number.

\subsection{Using the ``clash'' attack to reduce verification failures}

The following attack allows an attacker  who has intercepted many iVote sessions to share information between  them and hence  manipulate a large number of votes with limited detection.  The attack is a variant of the ``clash'' attack~\cite{kusters2012clash}.
% VT:  Actually it wasn't really introduced by Kusters et al.  It was folklore for a long time
% and just formalised by them.
We believe it would work, but of course we could not test it during the election without interfering with real votes.

When verification fails to produce the expected vote, the voter is supposed to complain to the authorities.  Inevitably, some voters will falsely complain, either mistakenly or  maliciously, that their correctly entered vote has been dropped or misrecorded.  The iVote verification design does not provide any evidence to support or disprove voter complaints, making it difficult to distinguish an  attack from the baseline level of complaints due to voter error.  This observation is important in the following attack, which reduces the number of complaints, but probably does not eliminate them altogether.
Although  this  attack would sometimes be detected, the percentage of verification   complaints would substantially underrepresent the fraction of manipulated votes, perhaps leading to an incorrect result being sustained during a post-election legal challenge.

The attack requires the ability to:
\begin{itemize}
        \item misdirect some voters' registrations,
        \item assign these voters a PIN at registration, as opposed to letting them choose, and
        \item compromise some iVote clients, using the attack from Section~\ref{sec:security} or simple misdirection.
\end{itemize}

First observe that, while the registration server itself was protected by HTTPS, the main iVote gateway from which voters reached it ran plain HTTP\footnote{Or rather, it did for the first week of voting, until we pointed this out to NSWEC.}\@.   This gave a man-in-the-middle attacker the opportunity to misdirect registration attempts to a site of the attacker's choosing, for instance by using the SSL\_strip attack~\cite{sslStrip}.  At this point the attacker could substitute a look-alike registration site with a modified workflow.  For instance, it could assign a PIN rather than accept one, under the assumption that a typical voter would not realize this was not the normal behavior.

Now note that Australian elections use  multi-candidate  preferential voting, so two voters who support party  $A$ may  subsequently list quite different lower preferences.  However,  some common patterns recur very often, for example the vote consisting  of a single (first) preference on each ballot.  Many voters also follow  official party ``How to Vote'' cards.  Although we are not aware of data  for NSW, studies in the neighboring state of Victoria show that overall  about 40\% of voters follow their how-to-vote card exactly~\cite{VECStatistics}.
    
The main idea of this attack is to intercept a voter's registration and give him the iVote ID and PIN of a like-minded person who has already voted, preferably one who has cast a simple vote likely to be repeated.  If the target voter's choices exactly match those of the first voter, then all of the verification will look exactly right to both voters.  The attacker can safely reuse the target voter's registration credentials to get a new iVote ID and PIN and cast an arbitrary vote.   If the target voter's choices are different from the first voter, he will detect a problem if he uses the verification service, but not if he contacts the receipt service only.

This attack removes a party-$A$ vote and substitutes a vote of the attacker's choice.  While it may sometimes be detected, if prediction of voter behavior is good then it raises far fewer complaints than that quantity of attacked votes ought to.  For  example, if prediction  is perfect then it raises no  alarm; if prediction is near-perfect then it manipulates many more votes than the number of verification complaints indicates.  Note that it is not hard to predict how someone will vote when you have their registration credentials and hence their electoral roll record.  

We find it notable that issues mentioned in the academic literature on verifiable voting---including the absence of dispute resolution  (or accountability) and the prospect of a particular kind of attack---here turn out to be relevant in the context of a real-world online election.

\section{Other Issues}
 
In this section we discuss additional problems and observations that we noted while assessing iVote.

\vspace{-6pt}
\subsection{Integrity, auditing, and verification}\label{sec:auditAndVerification}

The iVote verification and audit systems are incompletely described in public documents, and no source code is publicly available, so it is not possible for external independent observers like us to rule out the existence of other substantial risks to integrity, beyond those we have already described.  However, we can make several high-level observations about limitations of the design.

For instance, the design cannot achieve the same level of assurance for integrity as an ordinary post-election scrutineering process, since a related compromise of the Core Voting System and the Verification Service could undetectably alter votes.  For instance, the Verification Service could simply lie to the voter about what vote was recorded on their behalf.  Then the Core Voting System and Verification System could show consistent misrepresented votes to the Auditor.

The process for Auditing is incompletely described, so it is not clear whether a related compromise of the Core Voting System and the Auditor would also suffice to  alter votes undetectably.  A simple potential attack would be for the auditor to turn a blind eye to inconsistencies between the Core Voting System's data and the Verification Server's.  Would this be caught?  The Security Implementation Statement~\cite{ivoteSec} refers to some independent parties being allowed to observe some parts of the audit process and receive some software, but it does not say exactly what data they may check.
 
Votes that were present on the verification server (and possibly verified) could subsequently be removed if the voter re-registered or voted via another channel.  It is not clear from the published system description how or whether the auditor (or anyone else) could verify that only the correct votes were removed.

A compromised web server or Voice Server ({\it i.e.} the IVR system for phone voting) could perform the attack from Section~\ref{subsec:simpleAvoidance} on last-minute votes just as easily as a compromised web client.  This would be a low-risk attack, since the malicious server would know that the verification server would be turned off before the voter could perform verification and detect this.
% AH: I don’t understand the note below—the compromised server would just change the voter's input, submit a vote, and return the correct receipt number for the malicious vote, right?
%VT: There was in the JavaScript some sort of "signature" generated in the browser and sent in to the Voting Server.  Quite independent of receipt number etc.  I assume the PKI was established on the spot.  The point is whether the compromised server could leave the browser-side JS exactly as it was supposed to be and then just change the vote when it arrived, or whether the cheating server would have to change the JS served into the browser.  
\footnote{In the case of the web server, this would require forging a signature attached to the vote by the client.  This signing step is evident in the JavaScript, but we could not find any documentation on how the signing key was derived or how the signature was verified.  Hence we do not know whether a compromised web server could have simply created a new signature on any vote it received, or whether it would have needed to modify the JavaScript served to the client in order to get a valid signature on an altered vote.}
 
% AH: This doesn’t seem to fit without the footnote, and it’s mostly ironic anyway.  Proposing to strike.
%Registering required only the voter's name, date of birth and registered address.  A postcard-based notification system was designed to notify voters in the case that someone else registered fraudulently on their behalf.
%\footnote{Thanks to Barbara Simons for pointing out the irony of a mail-based security %mechanism in a system justified on the grounds of reaching those unable to access the post.}
 
% There are inconsistencies between different versions of NSWEC documents describing 
% the auditing process.   
% The ``iVote System Overview'' claims clearly (2.10.1) that the audit occurs ``without    
% revealing the voter preferences to the Auditor.'' 
% But this is    inconsistent with the Security   Implementation Statement, p. 12: ``An 
% independent auditor will check   that decrypted votes are matched to votes on the 
% Verification Service  to  ensure their validity.''  
 
There are important inconsistencies between the code and the documentation describing how votes are encrypted.  Early iVote documents~\cite{ivoteReports}, including The iVote System Overview, describe them as being encrypted with the Receipt Number; the Security Implementation Statement~\cite{ivoteSec} describes them as being  encrypted using the ElGamal public key encryption system with the public  keys of the Election and Verification Servers.  Our inspection of the JavaScript used by iVote clients indicates that neither description is completely accurate: votes are encrypted using a ``digital envelope''  which consists of a randomly-generated symmetric key, encrypted once each with the Election and Verification Servers' ElGamal public keys, plus the vote choices encrypted with AES  using the symmetric key.  This has implications for both the privacy and the integrity of the system.  Furthermore, the deviation of the actual code from the published specifications, particularly for such a central aspect of the voting protocol, raises 
broader questions about the accuracy of the published descriptions of iVote.
% goes beyond a simple lack of transparency, as it actively misleads external
% experts and the public.
%
% AH: I don't see any reason to be so kind to them, but we can put this back in if you want.
%This is a sensible way of encrypting data when taken in isolation (indeed, much more %sensible than the 10- or  12-digit symmetric keys that were described in earlier official %documents), but its deviation from the specification 

\subsection{Privacy}

The iVote approach of having voters telephone a third-party server to have their votes read back to them is unprecedented, either in Australia or (to our knowledge) elsewhere in the world. It introduces many different opportunities for privacy breaches and coercion after voting that do not exist in traditional paper-based voting.

For instance, a criminal could offer money in return for iVote verification credentials that produced the desired vote from the verification server, or a coercer could threaten punishment if such credentials are not provided.  As noted by McKay~\cite{mcKayCoercion}, such an attacker could use the Receipt Server to check that the voter had not revoted to change their selections.  Such attacks could originate anywhere in the world, and vote buying could even be automated---imagine a Tor hidden service that offered Bitcoin payments for proper votes.  

% AH: Propose cutting as too rhetorical: 
%It is one thing to acknowledge that remote voting is susceptible to coercion; it is quite another to add extra features that facilitate coercion.

Although the iVote design appears to give up on using technology to protect against vote buying and coercion, the system employs elaborate privacy measures to try to separate the voter’s identity from their ballot internally.  Encryption alone does not guarantee vote privacy, as the vote must eventually be counted somehow.  Some electronic voting systems, including the Norwegian Internet voting system \cite{gjosteen2012norwegian}, use verifiable mixing in order to hide the link between the decrypted vote and the encrypted form submitted by the voter.\footnote{Some also use homomorphic tallying, but that would not work for Australian (preferential) voting.} The ``cryptographic envelope'' form of encryption used in iVote does not seem conducive to these privacy-preserving tabulation methods. It is therefore crucial for privacy that the voter's identity cannot be reconnected with her symmetrically-encrypted vote, which seems to remain in the same recognizable form throughout the process. 

iVote tries to achieve this by storing various items of unique or private data in various different parts of the system, and the Security Implementation Statement~\cite{ivoteSec} makes reference to associations between these being destroyed.  However, compared to traditional postal ballots, for which the physical separation of the voter's identity from the ballot can occur irrevocably, the destruction of electronic links is much more difficult to achieve. This is especially true if components of the system are compromised or malfunctioning in ways that allows data to be observed, recorded, or transferred elsewhere.

Unfortunately, there are several critical places in the system where compromised components or malicious insiders could potentially associate voter identities with ballot contents.  For example:
 
\begin{enumerate}
\item  In the polling-site version of iVote, voters register and then vote via the same machine. This creates a single point of attack, as their identity and their vote are both present.
\item  All the phone communications, including voting and verifying by IVR system, are potentially susceptible to eavesdropping if the encryption used by the phone company is weak or absent.  This is particularly serious since both voting and verification involve transmitting the ballot contents over this channel, and since many voters use identifiable telephone numbers.
\item A compromise of the registration server, which knows the link between an individual's iVote ID and name, could be combined with only one other compromise (of the Verification Server, Voice Server, or possibly the Auditor) to link the name to the decrypted vote.
\item The verification server has simultaneous access to the voter's ballot contents and iVote ID.  If the voter accesses the service in a way that reveals their identity (for example, with a phone that has caller ID), then the verification server has all the information necessary to link the voter to their vote.  
%\Check{``Access to the  Verification Service will require the credentials  (iVote® Number and  PIN), plus the receipt number to be entered via the  telephone keypad.'',  Sec Imp Rep p\@.~35.   How hard is it to brute-force the hash and hence derive iVote Num   (+PIN), as well as Receipt Num and ciphertext? How are the iVote   num and PIN stored on the verification server?}
\end{enumerate}

\subsection{Usability and operations}

\begin{figure}[t]
    \centering
    \includegraphics[width=\linewidth]{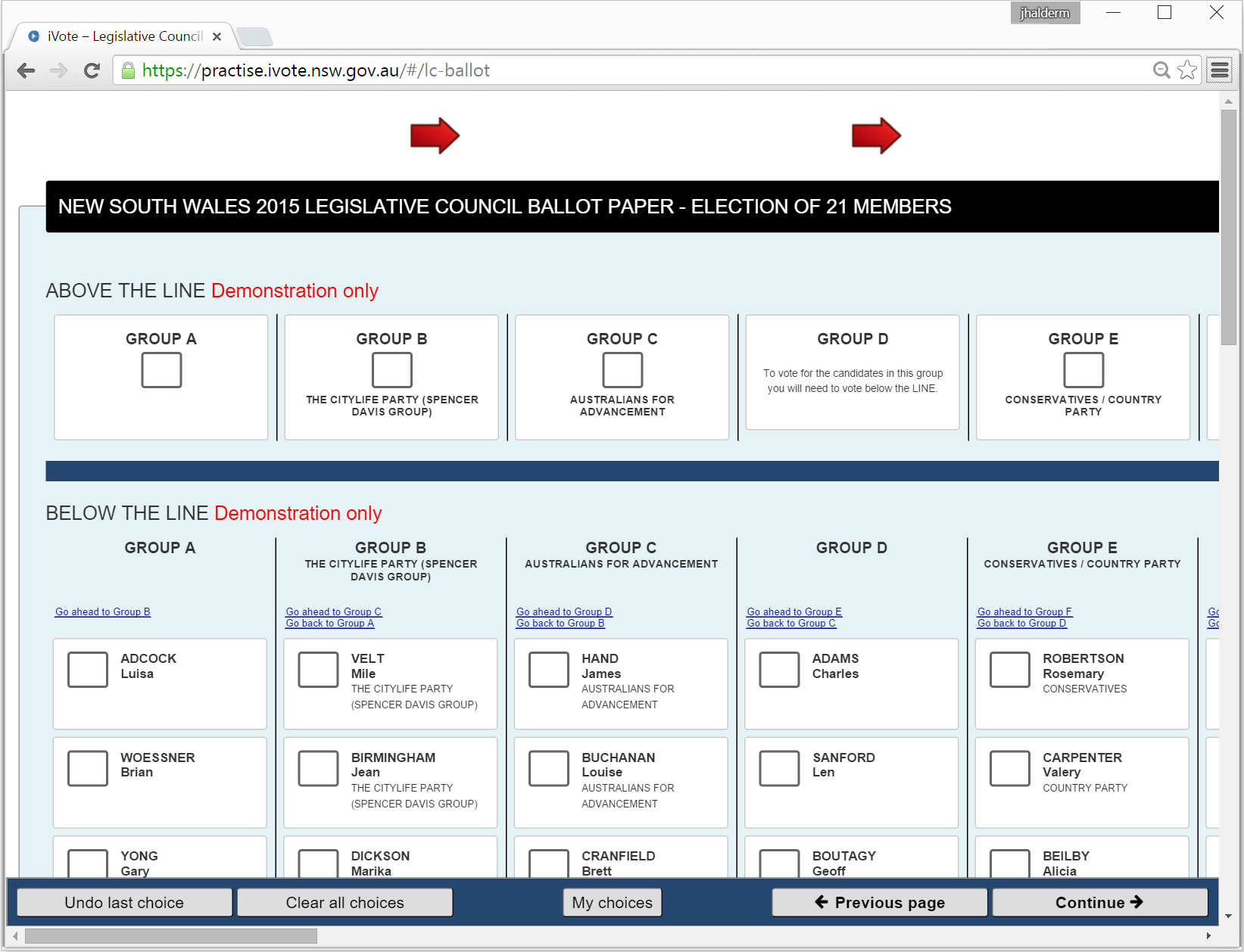}%ethertex: https://www.dropbox.com/s/rspjwn51ucpzy1d/usability-ballot.png?dl=1
     \caption{iVote suffered from problems beyond security.  Two parties  were mistakenly left off the ``above the line'' section of the ballot  for the first 19,000 votes, and the ballot interface (which required scrolling both horizontally and vertically to access all 394 candidates) was criticized for usability problems.\looseness=-1}
    \label{fig:UIArrows}
\end{figure}
 
iVote suffered other problems during the  election period.  The system was suspended for six hours because two minor parties had been left off the ``above the line'' section of the ballot.  The problem, blamed on human error, was fixed---but not before 19,000 votes had been cast.\looseness=-1
 
Other commentators (e.g.,~\cite{BadUI}) drew attention to serious usability problems with the ballot interface, which was very similar to the practice ballot design shown in Figure~\ref{fig:UIArrows}.  For instance, some voters reported difficulty navigating the ballot, which required scrolling horizontally and vertically to access all 24 party groups and 394 candidates. Scroll bars failed to appear on some browsers, and the red arrows at the top of the screen had no effect.  The ``Continue'' button with the right-pointing arrow ended the voting session and took the voter to a review, rather than advancing to the next part of the ballot as might be expected.
 
These problems would seem to suggest that the system's core voting functionality was not adequately tested prior to deployment.

\section{Lessons}

\subsection*{Security: the difficulty of correcting known problems in time, and unknown problems at all}

iVote's vulnerability to the FREAK and Logjam attacks illustrates once again why Internet voting is hard to do securely. The system had been in development for years, but FREAK was made public about two weeks before the election. New vulnerabilities are discovered regularly in software and protocols that an Internet voting system depends on for its security, including web browsers and TLS\@. When this happens near election day, there may not be time to ensure that election servers and voters' clients are properly retested and patched.

Moreover, mechanisms for trying to ensure that correct software is running in the voting system conflict with the necessity for rapid patching.  A last minute change to fix one serious problem could introduce new vulnerabilities---as happened in Washington, D.C.~\cite{dcFC}---or could conceal a deliberate attempt at fraud.\looseness=-1

The ability to test for and patch such problems assumes they are publicly known, but attackers may also have access to unpublished ``zero-day'' vulnerabilities for which, by definition, no patches yet exist.  This was the case for us with Logjam, which would have allowed us to compromise iVote connections to all popular browsers during the election. It is sheer luck that NSWEC's method of removing the vulnerability to FREAK also protected iVote from Logjam, as the attack was not disclosed publicly until two months later.  The only responsible assumption is that there are more major HTTPS vulnerabilities waiting to be discovered and perhaps already known to sophisticated attackers.
 
\vspace{-6pt}
\subsection*{Fragility: standard web development practices are inadequate for critical applications such as elections}
 
Many pieces of software contribute to a typical web application experience, including off-the-shelf server software and library code and, commonly, packages such as analytics tools that are loaded from third-party services.  While reliance on such components might be appropriate for a blog or even an e-commerce site, they are often not engineered to the level of security that is required for critical, high-risk applications.  (Indeed, analytics software has been shown to leak critical private information in certain online banking systems~\cite{pankkijuttu}.) Given the economic and foreign policy stakes involved in the outcome of a large election, such contests need to be treated as national security matters, which require a wholly different technical approach than typical IT systems.

Moreover, the decision to import code from a third party into the election system creates the possibility for that party to attempt to undermine the system.  Even if the PiwikPro server had not been vulnerable to man-in-the-middle attacks, anyone with administrative access to that server (whether legitimate or otherwise) would have been able to mount the same attack. Insider threats represent some of the most insidious security risks, and reliance on external code greatly expands the set of insiders who are able to affect the security of the election, adding possibly unknown employees of third-party service providers.

\vspace{-6pt}
\subsection*{Verifiability: when does an advertised verification mechanism truly provide verifiable evidence of a correct election outcome?}
 
Although some schemes do provide genuine electronic election verification remotely, including Helios \cite{helios}, Remotegrity~\cite{zagorski2013remotegrity}, and Pretty Good Democracy~\cite{ryan2013pretty}, achieving this in a privacy-preserving way requires real verification work from the voter.  Such techniques hold promise for the future, and have been used successfully in elections with relatively educated voters and low stakes~\cite{adida2009electing}.  However, extending these techniques to state-level elections remains impractical for now, and issues such as voter authentication and usability remain especially problematic. New South Wales is particularly challenging on these fronts, since it has no public key infrastructure and requires voters to number multiple preferences on a ballot with 394 candidates.

% AH: Propose cutting for space:
%There are sensible solutions for polling-place electronic election  verification, including voter-verifiable paper trails with paper counting or risk limiting audits~\cite{lindeman2012gentle}, and genuine end-to-end verifiable pollsite voting schemes~\cite{RyanBHSX09,bell2013star,carback2010scantegrity}.

Considering these limitations of state-of-the-art verification schemes, it is not surprising that the iVote verification mechanism was vulnerable to circumvention. It was not based on any peer-reviewed end-to-end verifiable scheme, and there was no detailed public review to allow such problems to be pointed out prior to the election. When an Internet voting system is claimed to be verifiable, this claim should be supported with a clear argument based on a complete description of the system. Otherwise the verification protocol itself could be incomplete, erroneous, or open to manipulation.  

\section{Conclusion}

We discovered serious flaws in the iVote online voting system that would have allowed a malicious attacker to expose voters' secret ballots, substitute replacement votes, and sidestep the verification mechanism. Despite years of planning, development, and pre-election security assessment, the system was susceptible to both publicly known and zero-day vulnerabilities that were at our disposal during the state election. These findings demonstrate yet again why conducting Internet voting with existing security technologies poses grave real-world risks.
 
NSWEC's decision to keep the system's source code and detailed design secret prevented independent analysts like us from being able to bring these specific problems to the officials' attention before the election.  Even now, we cannot know whether there are other critical flaws in the iVote software and protocols that would be evident if the relevant details were made public.
 
We recommend that NSWEC and others avoid large-scale Internet voting deployments until there are fundamental advances in computer security that can appropriately mitigate the risks.  If Internet voting tests must proceed, future tests should firmly restrict eligibility to voters unable to vote via a more secure channel; incorporate genuine, peer-reviewed verification mechanisms; ensure that the design and implementation are made openly available for rigorous independent scrutiny; and include a clear public statement of the risks to voter privacy and electoral integrity.\looseness=-1
 
Elections should produce not only an outcome but also sufficient evidence supporting that outcome.  This is the reason for Australia's tradition of transparent electoral processes, as well as for more recent research on auditable and verifiable elections.  In the case of the 2015 New South Wales state election, there is neither evidence that the vulnerabilities we discovered were exploited nor adequate proof that they were not. A demonstrable vulnerability exposing a large number of votes to potential manipulation constitutes a serious failure of the electoral process.

\ifblind\relax\else

\section*{Acknowledgments}

The authors thank David Adrian, Ed Felten, Rajeev Gor\'e, Nadia Heninger, Harri Hursti, and Liz Minchin for assistance during this project.  For their support and encouragement after we made our results public, we would also like to thank a tremendous community of election integrity scholars and advocates, including but not limited to: Duncan Buell, David Dill, Joseph Hall, Candice Hoke, David Jefferson, Noel Runyan, Ronald Rivest, Barbara Simons and Pamela Smith.

This material is based in part upon work supported by the U.S. National Science Foundation under grants CNS-1345254 and CNS-1409505,  and by the Morris Wellman Faculty Development Assistant Professorship. Any opinions, findings, and conclusions or recommendations expressed in this material are those of the authors and do not necessarily reflect the views of the National Science Foundation. % This disclaimer should be removed after peer review.

% by the Australian Research Council under grant DP140101119,  VT: This is a joint grant with
% Carroll, who would rather not sour the mood at NSWEC by appearing to criticise.  And anyway
% I didn't use any of the money from that grant for this project.
%
% AH: Fine with me, but if you want to disclose that funding, I suggest doing it here, rather than on the title page, since my equivalent disclosure is pretty long.

\fi

\bibliographystyle{abbrv}\urlstyle{rm}
\bibliography{paper}%ethertex: paper.bib=https://nsw.pad.jhalderm.com/bib

\begin{thebibliography}{10}

\bibitem{NSWECAssurance}
{ABC News}.
\newblock Computer voting may feature in {M}arch {NSW} election, Feb. 2015.
\newblock
  \url{http://www.abc.net.au/news/2015-02-04/computer-voting-may-feature-in-march-nsw-election/6068290}.

\bibitem{dnsbotnet}
O.~Abendan.
\newblock How {DNS} changer {T}rojans direct users to threats.
\newblock In {\em Trend Micro Threat Encyclopedia}, 2012.

\bibitem{helios}
B.~Adida.
\newblock Helios: {W}eb-based open-audit voting.
\newblock In {\em 17th USENIX Security Symposium}, Aug. 2008.
\newblock \url{https://vote.heliosvoting.org}.

\bibitem{adida2009electing}
B.~Adida, O.~De~Marneffe, O.~Pereira, and J.-J. Quisquater.
\newblock Electing a university president using open-audit voting: {A}nalysis
  of real-world use of {H}elios.
\newblock In {\em Electronic Voting Technology Workshop (EVT)}, 2009.

\bibitem{logJam}
D.~Adrian, K.~Bhargavan, Z.~Durumeric, P.~Gaudry, M.~Green, J.~A. Halderman,
  N.~Heninger, D.~Springall, E.~Thom\'{e}, L.~Valenta, B.~VanderSloot,
  E.~Wustrow, S.~Zanella-B\'{e}guelin, and P.~Zimmermann.
\newblock Imperfect forward secrecy: How {D}iffie-{H}ellman fails in practice,
  May 2015.
\newblock \url{https://weakdh.org/}.

\bibitem{bgphijacking}
H.~Ballani, P.~Francis, and X.~Zhang.
\newblock A study of prefix hijacking and interception in the {I}nternet.
\newblock In {\em Proceedings of ACM SIGCOMM}, Aug. 2007.

\bibitem{bell2013star}
S.~Bell, J.~Benaloh, M.~D. Byrne, D.~DeBeauvoir, B.~Eakin, G.~Fisher,
  P.~Kortum, N.~McBurnett, J.~Montoya, M.~Parker, et~al.
\newblock Star-vote: A secure, transparent, auditable, and reliable voting
  system.
\newblock {\em The USENIX Journal of Election Technology Systems, 1 (1)}, pages
  18--37, 2013.

\bibitem{beurdouche2015messy}
B.~Beurdouche, K.~Bhargavan, A.~Delignat-Lavaud, C.~Fournet, M.~Kohlweiss,
  A.~Pironti, P.-Y. Strub, and J.~K. Zinzindohoue.
\newblock A messy state of the union: {T}aming the composite state machines of
  {TLS}.
\newblock In {\em 36th IEEE Symposium on Security and Privacy}, 2015.

\bibitem{MooseMalware}
O.~Bilodeau and T.~Dupuy.
\newblock Dissecting {Linux/Moose}: The analysis of a {L}inux router-based worm
  hungry for social networks, May 2015.
\newblock
  \url{http://www.welivesecurity.com/wp-content/uploads/2015/05/Dissecting-LinuxMoose.pdf}.

\bibitem{carback2010scantegrity}
R.~Carback, D.~Chaum, J.~Clark, J.~Conway, A.~Essex, P.~S. Herrnson,
  T.~Mayberry, S.~Popoveniuc, R.~L. Rivest, E.~Shen, et~al.
\newblock Scantegrity {II} municipal election at {T}akoma {P}ark: The first
  {E2E} binding governmental election with ballot privacy.
\newblock In {\em Proceedings of the 19th USENIX conference on Security}, pages
  19--19. USENIX Association, 2010.

\bibitem{vvotePaper}
C.~Culnane, P.~Y.~A. Ryan, S.~Schneider, and V.~Teague.
\newblock {vVote}: A verifiable voting system.
\newblock {\em {ACM} Transactions on Information and System Security}.
\newblock To appear. Technical Report at \url{http://arxiv.org/abs/1404.6822}.

\bibitem{dambh:freak}
Z.~Durumeric, D.~Adrian, A.~Mirian, M.~Bailey, and J.~A. Halderman.
\newblock Tracking the {FREAK} attack.
\newblock \url{https://freakattack.com/}.

\bibitem{eestats}
{Estonian Internet Voting Committee}.
\newblock Statistics about {I}nternet voting in {E}stonia, May 2014.
\newblock
  \url{http://www.vvk.ee/voting-methods-in-estonia/engindex/statistics}.

\bibitem{gjosteen2012norwegian}
K.~Gj{\o}steen.
\newblock The norwegian internet voting protocol.
\newblock In {\em E-Voting and Identity}, pages 1--18. Springer, 2012.

\bibitem{NIST}
N.~Hastings, R.~Peralta, S.~Popoveniuc, and A.~Regenscheid.
\newblock Security considerations for remote electronic {UOCAVA} voting.
\newblock National Institute of Standards and Technology, NISTIR 7770, Feb.
  2011.
\newblock \url{http://www.nist.gov/itl/vote/upload/NISTIR-7700-feb2011.pdf}.

\bibitem{BadUI}
A.~Heber.
\newblock There's a huge design flaw in the {NSW} online voting system which
  {L}abor wouldn't be happy about.
\newblock Business Insider Australia, Mar.~28 2015.
\newblock
  \url{http://www.businessinsider.com.au/theres-a-huge-design-flaw-in-the-nsw-online-voting-system-which-labor-wouldnt-be-happy-about-2015-3}.

\bibitem{HeningerFactoring}
N.~Heninger.
\newblock Factoring as a service.
\newblock Crypto 2013 rump session.
\newblock \url{https://www.cis.upenn.edu/~nadiah/projects/faas/}.

\bibitem{dnspoisoning}
D.~Kaminsky.
\newblock It’s the end of the cache as we know it.
\newblock In {\em Toorcon}, 2008.

\bibitem{kusters2012clash}
R.~Kusters, T.~Truderung, and A.~Vogt.
\newblock Clash attacks on the verifiability of e-voting systems.
\newblock In {\em 33rd IEEE Symposium on Security and Privacy}, pages 395--409,
  2012.

\bibitem{sslStrip}
M.~Marlinspike.
\newblock New tricks for defeating {SSL} in practice.
\newblock Black Hat, 2009.
\newblock \url{http://www.thoughtcrime.org/software/sslstrip/}.

\bibitem{mcKayCoercion}
R.~McKay.
\newblock Flaws in {iVote's} re-vote process which attempts to defeat coercers.
\newblock \url{http://www.bigpulse.com/governmentelections#changevoteflaw}.
\newblock BigPulse.

\bibitem{ivoteReports}
{NSW Electoral Commission}.
\newblock Index of {iVote} reports.
\newblock
  \url{http://www.elections.nsw.gov.au/about_us/plans_and_reports/ivote_reports}.

\bibitem{ivoteFAQ}
{NSW Electoral Commission}.
\newblock {iVote}: {F}requently asked questions.
\newblock \url{https://www.ivote.nsw.gov.au/faq.aspx}.

\bibitem{NSW2015LCResults}
{NSW Electoral Commission}.
\newblock {NSW} 2015 legislative council election - final distribution of
  preferences.

\bibitem{ivoteThreats}
{NSW Electoral Commission}.
\newblock {iVote} threat analysis and risk assessment, Jan. 2014.
\newblock
  \url{http://www.elections.nsw.gov.au/__data/assets/pdf_file/0008/175760/NSW_Election_-_iVote_Threat_Analysis_and_Risk_Assessment_v3.0.pdf}.

\bibitem{ivoteSec}
{NSW Electoral Commission}.
\newblock {iVote} system security implementation statement, Mar. 2015.
\newblock
  \url{http://www.elections.nsw.gov.au/__data/assets/pdf_file/0007/193219/iVote-Security_Implementation_Statement-Mar2015.pdf}.

\bibitem{pankkijuttu}
O.~R\"ais\"anen.
\newblock The bank deal.
\newblock \url{http://oona.windytan.com/pankki.html}.

\bibitem{ryan2013pretty}
P.~Y. Ryan and V.~Teague.
\newblock Pretty good democracy.
\newblock In {\em Security Protocols XVII}, pages 111--130. Springer, 2013.

\bibitem{nostats}
B.~Segaard, D.~A. Christensen, B.~Folkestad, and J.~Saglie.
\newblock Internettvalg: {H}va gj{\o}r og mener velgerne?, 2014.
\newblock
  \url{https://www.regjeringen.no/globalassets/upload/kmd/komm/rapporter/isf_internettvalg.pdf}.

\bibitem{estoniaCCS}
D.~Springall, T.~Finkenauer, Z.~Durumeric, J.~Kitcat, H.~Hursti, M.~MacAlpine,
  and J.~A. Halderman.
\newblock Security analysis of the {E}stonian {I}nternet voting system.
\newblock In {\em ACM Conference on Computer and Communications Security
  (CCS)}, Nov. 2014.

\bibitem{FreedomToTinker}
V.~Teague and J.~A. Halderman.
\newblock Security flaw in {N}ew {S}outh {W}ales puts thousands of online votes
  at risk.
\newblock Freedom to Tinker blog post, Mar.~22 2015.
\newblock
  \url{https://freedom-to-tinker.com/blog/teaguehalderman/ivote-vulnerability/}.

\bibitem{TheConversation}
V.~Teague and J.~A. Halderman.
\newblock Thousands of {NSW} election online votes open to tampering.
\newblock The Conversation, Mar.\ 23 2015.
\newblock
  \url{https://theconversation.com/thousands-of-nsw-election-online-votes-open-to-tampering-39164}.

\bibitem{VECStatistics}
{Victorian Electoral Commission}.
\newblock Report to {P}arliament on the 2010 {V}ictorian {S}tate election;
  {S}ection 11: {S}tatistical overview of the election, 2011.
\newblock \url{http://www.vec.vic.gov.au/files/ER-2010-Section11.pdf}.

\bibitem{dcFC}
S.~Wolchok, E.~Wustrow, D.~Isabel, and J.~A. Halderman.
\newblock Attacking the {W}ashington, {D}.{C}. {I}nternet voting system.
\newblock In {\em 16th International Conference on Financial Cryptography and
  Data Security (FC)}, Feb. 2012.

\bibitem{zagorski2013remotegrity}
F.~Zag{\'o}rski, R.~T. Carback, D.~Chaum, J.~Clark, A.~Essex, and P.~L. Vora.
\newblock Remotegrity: {D}esign and use of an end-to-end verifiable remote
  voting system.
\newblock In {\em Applied Cryptography and Network Security (ACNS)}, pages
  441--457. Springer, 2013.

\end{thebibliography}

\end{document}